\documentclass[aps,prl,twocolumn,showpacs,superscriptaddress]{revtex4}
\usepackage{graphicx}
\usepackage{epsfig}
\usepackage{microtype}
\usepackage[dvips]{color}

\begin{document}
\bibliographystyle{prsty}

\title{Entanglement of macroscopic test masses and \\ the Standard Quantum Limit in
       laser interferometry}

\author{Helge M\"uller-Ebhardt}
\affiliation{Max-Planck-Institut f\"ur Gravitationsphysik
(Albert-Einstein-Institut), \\ Institut f\"ur Gravitationsphysik,
Leibniz Universit\"at Hannover, Callinstr. 38, 30167 Hannover,
Germany}
\author{Henning Rehbein}
\affiliation{Max-Planck-Institut f\"ur Gravitationsphysik
(Albert-Einstein-Institut), \\ Institut f\"ur Gravitationsphysik,
Leibniz Universit\"at Hannover, Callinstr. 38, 30167 Hannover,
Germany}
\author{Roman Schnabel}
\affiliation{Max-Planck-Institut f\"ur Gravitationsphysik
(Albert-Einstein-Institut), \\ Institut f\"ur Gravitationsphysik,
Leibniz Universit\"at Hannover, Callinstr. 38, 30167 Hannover,
Germany}
\author{Karsten Danzmann}
\affiliation{Max-Planck-Institut f\"ur Gravitationsphysik
(Albert-Einstein-Institut), \\ Institut f\"ur Gravitationsphysik,
Leibniz Universit\"at Hannover, Callinstr. 38, 30167 Hannover,
Germany}
\author{Yanbei Chen}
\affiliation{Max-Planck-Institut f\"ur Gravitationsphysik
(Albert-Einstein-Institut), Am M\"uhlenberg 1, 14476 Potsdam,
Germany}

\date{\today}

\begin{abstract}
We show that the generation of entanglement of two heavily
macroscopic mirrors is feasible with state of the art techniques
of high-precision laser interferometry. The basis of such a
demonstration would be a Michelson interferometer with suspended
mirrors and simultaneous homodyne detections at both
interferometer output ports. We present the connection between the
generation of entanglement and the Standard Quantum Limit (SQL)
for a free mass. The SQL is a well-known reference limit in
operating interferometers for gravitational-wave detection and
provides a measure of when macroscopic entanglement can be
observed in the presence of realistic decoherence processes.
\end{abstract}


\maketitle

The continuous quantum measurement of macroscopic objects was
first investigated in the context of gravitational-wave (GW)
detection~\cite{SQL}. In laser interferometer GW detectors,
incoming GWs induce very weak, yet highly classical, tidal forces
on mirror-endowed test masses. These are approximately free masses
since they are hung from seismic isolation stacks as pendulums
with eigenfrequency ($\sim 1$\,Hz) much below detection band
(10\,Hz $\lesssim \Omega/(2\pi) \lesssim$ 10\,kHz). The
interferometer measures the change in the test-mass mirror's
relative positions~\cite{Abr1992}. The position observable,
however, does not have commuting Heisenberg operators at different
times, resulting in  a Standard Quantum Limit (SQL)~\cite{SQL}
\begin{equation}
\label{SQL} S_{\rm SQL}  \equiv  2\hbar/(m\Omega^2),
\end{equation}
where $m$ is the reduced mirror mass. After a rather controversial
debate it was eventually realized that the SQL can in principle be
surpassed~\cite{QND} if (i) measurement of the external force is
made through monitoring of a {\it quantum non-demolition}
observable~\cite{BrKh1996}, (ii) the quantum mechanics of the
measuring device is taken into account, and appropriate quantum
correlations are used, e.g., with {\it back-action evasion}
techniques~\cite{ThorneCavesZimmermann1978}. To reach the
sensitivity required for GW astronomy, future GW detectors as well
as near term prototypes will be designed to reach and even surpass
the SQL~\cite{Fri2003,SQUEEZER,Miy2004,KLMTV2001,AdLIGO} using
either of the approaches mentioned.  At the same time it is
necessary to reduce all  technical noise in order to reach this
goal. These efforts are made with devices having test masses and
arms ranging from gram-scale and meter-long~\cite{SQUEEZER},
kilogram-scale and 100\,m~\cite{Miy2004} up to 40\,kg and
4\,km~\cite{Fri2003,KLMTV2001,AdLIGO}.

In this Letter, we show that a Michelson interferometer at its SQL
produces entanglement between the {\it conditional states} of its
otherwise free test-mass mirrors, if both common- and
differential-mode test-mass motion are measured, but with
different quantum-measurement processes -- a direct mechanical
analogy to the Einstein-Podolsky-Rosen gedanken
experiment~\cite{EPR1935}. Contrary to former
proposals~\cite{entanglements}, our work is aimed directly at
experimental efforts in the GW community towards SQL sensitivity,
and therefore may enable the first observation of entanglement
between truly macroscopic objects.  In technical terms, we
introduce {\it Wiener Filtering}~\cite{Wiener1949}, which,
compared with the more conventional approach involving Stochastic
Master Equation (SME)~\cite{Mil1996}, allows us to determine the
conditional state much more directly from the interferometer's
noise budget.
\begin{figure}[ht]
\centerline{\includegraphics[width=6.8cm]{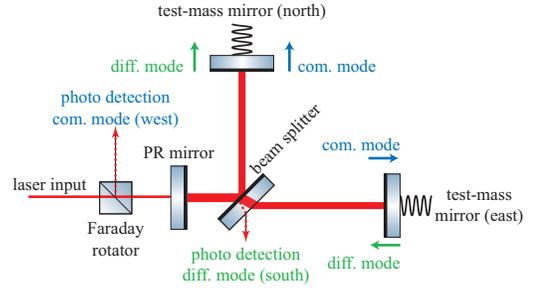}}
\caption{Schematic plot of a power-recycled (PR) Michelson
interferometer. Suspended test-mass mirrors are much lighter than
other suspended optics. Differential motion of test-mass mirrors
detected at dark (south) port and common motion at bright (west)
port. A Faraday rotator might be used to access all of the back
reflected light.\label{mich}}
\end{figure}

{\it Configuration.} We consider an equal-arm Michelson
interferometer as in Fig.\,\ref{mich}, with laser (carrier) light
injected from left. The two beams split at the beam splitter (BS)
are reflected by identical mirrors in the north (n) and east (e)
arms, before being recombined at the BS.  The south port is kept
dark at the zero point, with all light reflected to the bright
port. A power-recycling (PR) mirror is positioned such that it
forms a resonant, but relatively low finesse, cavity for the
carrier light with the test-mass mirrors. A differential (common)
arm-length change induces phase-modulation fields that only emerge
at the dark (bright) port. Correspondingly, fluctuating modulation
fields that enter the interferometer from the dark (bright) port
only interact with test-mass mirror displacements in the
differential (common) mode. Homodyne detections are made at both,
the dark and the bright port, each with a certain
frequency-independent quadrature phase. In this way, we have two
independent measurement processes in our interferometer, one for
$\hat x^{\rm d} \equiv (\hat x^{\rm e} - \hat x^{\rm n})$ and the
other for $\hat x^{\rm c} \equiv (\hat x^{\rm e} + \hat x^{\rm
n})$. Defining  $\hat p^{\rm c, d} \equiv (\hat p^{\rm e} \pm \hat
p^{\rm n})/2$,  we have $[\hat x^{\rm c,d},\hat p^{\rm c,d}]=
i\hbar$. As it is well known, homodyne detections of out-going
modulation fields collapse the quantum state of the corresponding
mode of test-mass mirror motion. In absence of classical noise,
each mode will  reach a stochastic Gaussian pure state, with
first-order moments $\langle \hat x \rangle$ and $\langle \hat
p\rangle$ undergoing random walk, and second-order moments
$V_{xx}$, $V_{xp}$ and $V_{pp}$ remaining constant and
minimum-Heisenberg-limited, as required by purity. At any instant,
values of $\langle \hat x(t) \rangle$ and $\langle \hat p(t)
\rangle$ are determined by measurement results in the past. In
this sense, this Gaussian state is called a {\it posterior
state}~\cite{Barchielli1992}. As long as the measurement processes
for the common and differential modes are different in terms of
e.g. signal storage time and/or homodyne phase, the wave functions
$\psi^{\rm c}$ and $\psi^{\rm d}$ will be different, and the joint
wave function $\Psi(x^{\rm e}, x^{\rm n}) = \psi^{\rm c}[(x^{\rm
e} + x^{\rm n})/2] \ \psi^{\rm d}[(x^{\rm e} - x^{\rm n})/2]$ must
be non-separable. However, in presence of classical noise
entanglement becomes less significant or even disappears. Note
that this is analogous to creating entanglement by overlapping two
differently squeezed beams on a BS~\cite{OPTIC}.

{\it Wiener Filtering.} In the following we obtain the conditional
quantum state using Wiener Filtering~\cite{Wiener1949}, which
applies to stable linear systems with Gaussian noise. More details
will be given in Ref.\,\cite{Che2007}. Let $\hat{y}(t)$ be the
Heisenberg operator of the out-going field quadrature being
measured, and $\hat{x}(t)$ be the Heisenberg operator of the
test-mass position. If we want to estimate $\hat x$ using the
output $\hat y$, the optimal filter $K_x(t)$ is determined by the
condition that the quantity
\begin{equation} \label{filter}
\hat R_x (t) \equiv \hat x (t) - \int_{-\infty}^{t} {\rm d} t' K_{x}
(t-t') \ \hat y (t')
\end{equation}
must be uncorrelated with the past output, i.e. $\langle \hat R_x
(t) \hat y (t')  + \hat y (t') \hat R_x (t)\rangle = 0$ for all
$t'<t$. This leads to a Wiener-Hopf equation, whose solution can
be given in terms of the single-sided (cross-) spectral densities
of $\hat x$ and $\hat y$: $S_{xy}(\Omega)$, $S_{x}(\Omega)$ and
$S_{y}(\Omega) \equiv s_y (\Omega)s_y^*(\Omega)$ with
$s_y(\Omega)$ and its inverse being analytic in the upper-half
complex plane. Then $K_x(\Omega) =
\left[S_{xy}/s_{y}^*\right]_+/s_{y}\,.$ Note that $[\ldots]_+$
stands for taking the component of a function whose inverse
Fourier transform has support only in positive times. The first-
and second-order moments of the posterior state are then given by
\begin{eqnarray}
\langle \hat x(t) \rangle &=& \int_{-\infty}^t  dt' K_x(t-t') y (t'), \\
V_{xx} &=& \int_0^\infty \frac{d\Omega}{2\pi}\left(S_{x}-
\left[{S_{xy}}/{s_{y}^*}\right]_+\left[{S_{xy}}/{s_{y}^*}\right]_+^*\right),
\label{wienervar}
\end{eqnarray}
respectively, where $y (t)$ is the measurement data of
$\hat{y}(t)$. The same approach applies to the momentum operator
$\hat p$ (replace every $x$ by $p$ in
Eqs.\,(\ref{filter})--(\ref{wienervar})), and we further find
\begin{equation}
V_{xp} = \int_0^\infty \frac{d\Omega}{2\pi}\ \Re\left\{S_{xp}-
\left[S_{xy}/s_{y}^*\right]_+\left[S_{py}/s_{y}^*
\right]_+^*\right\}.
\end{equation}
Such an essentially classical filtering is justified because
$[\hat y(t), \hat y(t')]=0$ (satisfied by all out-going field
quadratures) and $[\hat x(t), \hat y(t')]=[\hat p(t), \hat
y(t')]=0$ for $t>t'$ (due to causality, i.e. the out-going field
does not influence any future test-mass observable). Furthermore,
it can be shown that the conditional moments do not change when
different linear feedback systems are applied -- as long as at
time $t$ the control force only depends on $\{\hat y(t'): t'<t\}$.

{\it Posterior state of a single mode and squeezing.} Both, the
common and the differential mode quantum measurement process, can
each be described by a set of two frequency-domain Heisenberg
equations,
\begin{eqnarray}
\hat y (\Omega) &=&\sin \phi \ \hat a_1 + \cos \phi  \left[\hat a_2
+ {\alpha}/{\hbar} (\hat x (\Omega) + \hat \xi_x)\right],
\label{eomy}\\
\hat x (\Omega) &=& -  (\alpha \ \hat a_{1} + \hat \xi_F)/\left[m
(\Omega^2 +  i\gamma_m\Omega-\omega_m^2)\right], \label{eomx}
\end{eqnarray}
with $\omega_m$ the pendulum angular frequency and $\gamma_m$ the
pendulum damping rate. Eq.\,(\ref{eomy}) shows that $\hat y$
depends on the in-going amplitude/phase quadratures $\hat
a_{1,2}$, the motion of the mirror center of mass, and {\it
sensing noise} $\hat \xi_x$. The latter may be due to optical
losses, or thermal fluctuations of the mirror's shape, i.e.
internal thermal noise, which makes the mirror surface move with
respect to its center of mass. Eq.\,(\ref{eomx}) describes the
motion of the test mass center of mass under radiation-pressure
noise ($\alpha \ \hat a_1$), as well as a classical {\it force
noise} $\hat{\xi}_F$, acting directly on the center of mass of
each test-mass mirror, e.g. due to seismic noise or suspension
thermal noise. The quantity $\alpha$ indicates the measurement
strength, given by $\alpha_{\rm d}=\sqrt{4 \hbar\omega_0 P}/c$ for
the differential mode and $\alpha_{\rm c} = 2/\tau \sqrt{4
\hbar\omega_0 P}/c$ for the common mode, where $P$ is the
circulating power in the arms and $\tau$ the PR-mirror
transmissivity. The angle $\phi$ denotes the quadrature angle of
$\hat y$. In the following we will only consider phase-quadrature
readout, i.e. $\phi=0$.

\begin{figure}[t]
\centerline{\includegraphics[height=8cm,angle=-90]{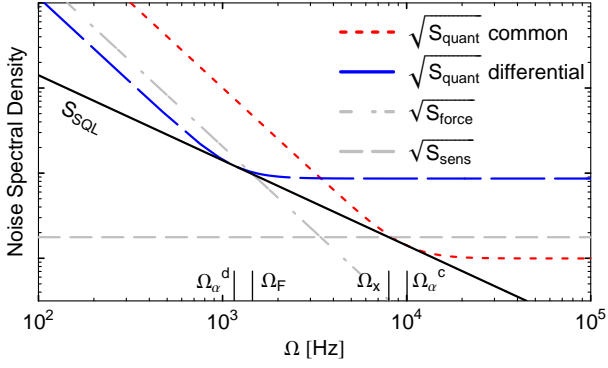}}
\caption{Noise spectral densities in arbitrary units with
$\Omega_F = 2 \pi$\,230\,Hz and $\Omega_x = 2 \pi$\,1270\,Hz, and
optimized $\Omega_\alpha^{\rm c} = 2 \pi$\,1600\,Hz,
$\Omega_\alpha^{\rm d} = 2 \pi$\,184\,Hz. We obtain a test-mass
entanglement of $E_{\mathcal N} = 0.35$. \label{snd}}
\end{figure}

From Eqs.\,(\ref{eomy}),(\ref{eomx}) and with $S_{a_i a_j} =
\delta_{ij}$ (referring to vacuum input) as well as $S_{\xi_F}$
and $S_{\xi_x}$, determined from the interferometer's noise
budget, we obtain the relevant spectral densities: $S_x =
[\alpha^2 +
S_{\xi_F}]/[m^2((\Omega^2-\omega_m^2)^2+\gamma_m^2\Omega^2)]$,
$S_{xy} = (\alpha/\hbar) S_x$ and $ S_{y}=1+(\alpha/\hbar)^2 [S_x
+ S_{\xi_x}]$ as well as $S_p = m^2\Omega^4 S_x$, $S_{py}=-{\rm
i}m\Omega S_{xy}$ and $S_{xp} = {\rm i}m\Omega S_x$. For
simplicity, we assume $\hat \xi_F$ and $\hat \xi_x$ to have a
white spectrum, $S_{\xi_F} = 2\hbar m \Omega_F^2$ and $S_{\xi_x} =
2\hbar/( m \Omega_x^2 )$, even though our approach enables us to
treat non-white spectra. The sensitivity of the interferometer is
given by the SQL-normalized, position-referred noise spectral
density in the detection band, i.e. $\Omega \gg \omega_m,\gamma_m$
\begin{equation}
\frac{S_{\rm total}}{S_{\rm SQL}} =\underbrace{
\frac{1}{2}\left(\frac{\Omega^2}{\Omega_\alpha^2}+\frac{\Omega_\alpha^2}{\Omega^2}\right)}_{S_{\rm
quant}/S_{\rm SQL}} + \underbrace{\Omega_F^2/\Omega^2}_{S_{\rm
force}/S_{\rm SQL}} + \underbrace{\Omega^2/\Omega_x^2}_{S_{\rm
sens}/S_{\rm SQL}}.
\end{equation}
The quantum noise $S_{\rm quant}$ is limited from below by $S_{\rm
SQL}$ with equality reached at $\Omega_\alpha \equiv
\alpha/\sqrt{m \hbar}$ which sets a time scale for the measurement
process. The force and sensing noise $S_{\rm force}$ and $S_{\rm
sens}$ intersect $S_{\rm SQL}$ at $\Omega_F $ and $\Omega_x$,
respectively.  For $\Omega_x/\Omega_F > 2$, we have a non-zero
frequency band (in between $\Omega_F$ and $\Omega_x$) in which the
classical noise is completely below the SQL (cf. Fig.\,\ref{snd}).
Solving the Wiener Filtering problem, we obtain in the limit of
$\gamma_m$, $\omega_m \ll \Omega_{\alpha}$
\begin{eqnarray}
\label{condvarxx}
V_{xx} &=&
\hbar/(\sqrt{2} m\Omega_\alpha)
{[(1+2\zeta_F^2)(1+2\zeta_x^2)^3]}^{1/4},\\
V_{pp} &=&\hbar m \Omega_\alpha/\sqrt{2}
{[(1+2\zeta_F^2)^3(1+2\zeta_x^2)]}^{1/4},\\
\label{condvarxp} V_{xp}&=&\hbar/2
[(1+2\zeta_F^2)(1+2\zeta_x^2)]^{1/2},
\end{eqnarray}
for the posterior state, where we have defined $\zeta_F\equiv
\Omega_F/\Omega_\alpha$ and $\zeta_x \equiv
\Omega_\alpha/\Omega_x$. At the bright port the in-going
modulation fields at frequencies $\Omega$ are usually not in
vacuum states, i.e. $S_{a_{1}a_{1}},S_{a_{2}a_{2}} > 1$, due to
technical laser noise, which adds for the common mode
$(S_{a_{1}a_{1}} -1)/2$ and $(S_{a_{2}a_{2}} -1)/2$ to
$\zeta_{F}^2$ and $\zeta_{x}^2$, respectively. Note that
Eqs.\,(\ref{condvarxx})--(\ref{condvarxp}) are consistent with
results obtained from SME~\cite{HJHS2003}, although with a more
complicated noise budget, solving the Wiener Filtering problem
will be substantially easier.

\begin{figure}[h!!!]
\centerline{\includegraphics[height=6cm,angle=-90]{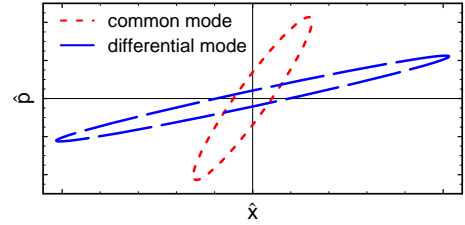}}
\caption{Squeezing ellipses of position and momentum operators
with respect to common and differential mode for the same
parameters as in Fig.\,\ref{snd}. All second-order moments are
normalized to the ground state of an harmonic oscillator at
frequency $(\Omega_\alpha^{\rm c} + \Omega_\alpha^{\rm d})/2$.}
\label{ellipse}
\end{figure}

A measure of the quantum nature of a state is the quantity $U =
V_{xx} V_{pp} - V_{xp}^2$, with $U \ge \hbar^2/4$, the Heisenberg
Uncertainty Principle. According to
Eqs.~(\ref{condvarxx})--(\ref{condvarxp}),
\begin{equation}
{U}/{(\hbar^2/4)} =(1+2\zeta_F^2)(1+2\zeta_x^2) \ge
(1+2\Omega_F/\Omega_x)^2,
\end{equation}
with equality achieved when $\Omega_\alpha = \sqrt{\Omega_x
\Omega_F}$. We see that as long as $\zeta_F$ and $\zeta_x$ are not
too big, the posterior state is highly squeezed in position, and
highly anti-squeezed in momentum with respect to the ground state
of the pendulum (recall that $\Omega_\alpha\gg\omega_m$).
Increasing (decreasing) $\Omega_{\alpha}$, e.g. by varying the
circulating power or the PR-mirror transmissivity, results in more
(less) squeezing in position, and more (less) anti-squeezing in
momentum -- causing the squeezing ellipse to rotate, as shown in
Fig.\,\ref{ellipse}. The posterior state can also be modified when
a different output-quadrature phase $\phi$ is used.

\begin{figure}[h!!!]
\centerline{\includegraphics[height=8cm,angle=-90]{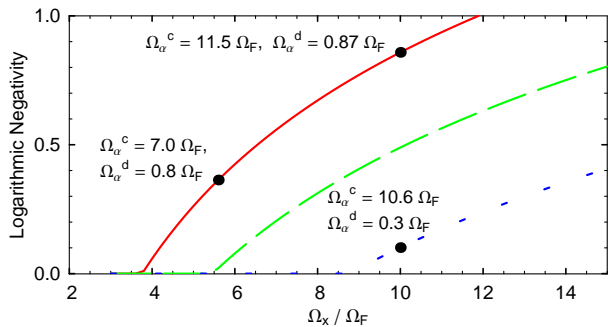}}
\caption{Logarithmic negativity versus $\Omega_x/\Omega_F$,
maximized with respect to $\Omega_\alpha^{\rm c}$ and
$\Omega_\alpha^{\rm d}$ (example values marked) for no technical
laser noise (solid line), 5\,dB (dashed line) and 10\,dB (dotted
line) technical laser noise, in both the amplitude and phase
quadratures.} \label{LogNeg2}
\end{figure}

{\it Entanglement.} After obtaining the individual common- and
differential-mode posterior states, we assemble the posterior
state of the entire system. The combined ($4 \times 4$) covariance
among $(x^{\rm e}, p^{\rm e}, x^{\rm n}, p^{\rm n})$ reads $V =
\left(\left(V_{\rm ee}, V_{\rm en} \right), \left(V_{\rm ne},
V_{\rm nn}\right)\right)$ with
\begin{eqnarray}
V_{\rm nn}=V_{\rm ee}&=&  \left(\begin{array}{cc} (V_{xx}^{\rm c}
+ V_{xx}^{\rm d})/4
& (V_{xp}^{\rm c} + V_{xp}^{\rm d})/2 \\
(V_{xp}^{\rm c} + V_{xp}^{\rm d})/2 & V_{pp}^{\rm c} + V_{pp}^{\rm
d} \\ \end{array} \right), \\
V_{\rm en}=V_{\rm ne}&=&  \left(\begin{array}{cc} (V_{xx}^{\rm c}
- V_{xx}^{\rm d})/4 & (V_{xp}^{\rm c} - V_{xp}^{\rm d})/2
\\ (V_{xp}^{\rm c} - V_{xp}^{\rm d})/2 & V_{pp}^{\rm
c} - V_{pp}^{\rm d}\\ \end{array} \right).
\end{eqnarray}
A computable measure of entanglement, the {\it logarithmic
negativity}, for an arbitrary bipartite system was introduced
in~\cite{ViWe2002}. For our state
\begin{equation} \label{logneg}
E_{\mathcal N} = \max[0, - \log_2 2 \sigma^-/\hbar],
\end{equation}
where $\sigma^- = \sqrt{(\Sigma - \sqrt{\Sigma^2 - 4 \det V})/2}$
and $\Sigma = \det V_{\rm nn} + \det V_{\rm ee} - 2 \det V_{\rm
ne}$. If we insert Eqs.\,(\ref{condvarxx})--(\ref{condvarxp}) into
Eq.\,(\ref{logneg}), only three parameter ratios remain:
$\Omega_\alpha^{\rm c,d}/\Omega_F$ for common and differential
mode, respectively, and $\Omega_x/\Omega_F$. The latter turns out
to be the crucial factor for the existence of test-mass
entanglement. The parameters in Fig.\,\ref{snd} result in
$E_{\mathcal N} = 0.35$. Recall that there exists a frequency
window with sub-SQL classical noise iff $\Omega_x/\Omega_F>2$.
However, the existence of entanglement sets a slightly higher
threshold value for this frequency ratio. In Fig.\,\ref{LogNeg2},
we plot the maximum achievable $E_{\cal{N}}$ (varying
$\Omega_{\alpha}^{\rm c}$ and $\Omega_{\alpha}^{\rm d}$) as a
function of $\Omega_x/\Omega_F$ for different strengths of
technical laser noise. Remarkably, even for a value of 10\,dB
above vacuum noise, the generation of entanglement is still
possible.

{\it Feasibility of achieving sub-SQL classical noise.} Current
experiments~\cite{Fri2003,SQUEEZER,Miy2004,KLMTV2001,AdLIGO} aim
to reach or even surpass the SQL by reducing the classical noise
around $\Omega_\alpha$ which is in fact not the optimum strategy
for strong test-mass entanglement (cf. Fig.\,\ref{snd}).
Furthermore, these experiments only consider a single readout.
Nevertheless, they provide valuable information whether sub-SQL
classical noise is achievable at all. Regarding the force noise,
it is feasible to suppress the dominant suspension thermal noise
below the SQL in the detection band even at room temperature. This
can be realized by a low pendulum frequency with a high quality
factor $\omega_m/\gamma_m$. In Advanced LIGO~\cite{AdLIGO}, we
expect $\Omega_F/(2\pi) \sim 40$\,Hz and $\Omega_\alpha/(2\pi)$
marginally higher. In lab-scale experiments $\Omega_F/(2\pi) =
400\,$Hz and $\Omega_\alpha/(2\pi)$ of several kHz have already
been realized~\cite{SQUEEZER}. Regarding the sensing noise,
optical losses need to be discussed~\cite{KLMTV2001}. In our setup
the main contribution arises from photodetection loss because no
high-finesse cavities are involved. Assuming a total optical loss
in power of $\epsilon = 2\%$, this on its own would imply
$\Omega_x =14 \Omega_\alpha$, which is more than sufficient. More
challenging is mirror internal thermal noise. In Advanced LIGO, we
expect $\Omega_x$ to roughly coincide with $\Omega_\alpha$.
However, cooled test masses will allow to suppress this noise and
increase $\Omega_x$. In the CLIO interferometer currently being
commissioned, the theoretical thermal-noise budget suggests
$\Omega_x/\Omega_F \sim 10$ at a temperature of
20\,K~\cite{Miy2004}. In the configuration outlined in
Ref.\,\cite{SQUEEZER}, we have $\Omega_x/(2\pi)$=50\,Hz, which is
a factor of 8 below $\Omega_F$. With a new test-mass with lower
mechanical coating loss, we expect $\Omega_x$ to increase by a
factor of $\sim 1.5$. Increasing the beam spot size by factor of
3, will provide an additional factor of $3^2$~\cite{SQUEEZER},
allowing $\Omega_x/\Omega_F\sim 1.7$, even at room temperature.
Other strategies are currently being investigated to further
suppress the internal thermal noise~\cite{coating}. Laser {\it
amplitude} noise can be suppressed sufficiently with current
technology~\cite{Seif2006}. Regarding laser {\it phase} noise, a
level of 10\,dB above vacuum noise -- for an input laser power of
e.g. 1\,W at a measurement frequency of 1\,kHz -- corresponds to a
frequency noise of about $10^{-5}\,{\rm Hz}/\sqrt{{\rm Hz}}$.
Ref.\,\cite{Camp2000} suggests that this stability has already
been reached for the injected carrier light of current GW
detectors.

{\it Conclusion.} In this Letter, we obtain a general and
quantitative condition for the possibility of generating
macroscopic entanglement between two test-mass mirrors of a
Michelson interferometer -- in terms of the interferometer's
classical noise. We have found that these classical noise sources
are required to surpass the SQL only moderately, well within the
reach for experiments to be carried out in the GW community within
the next decade.

\begin{acknowledgments}
{\it Acknowledgments.} We thank the AEI-Caltech-MIT MQM  group for
interesting discussions, in particular, T.~Corbitt, C.~Li,
N.~Mavalvala, Y.~Mino, K.~Somiya, K.S.~Thorne, S.~Waldman, and
C.~Wipf.  We thank B.~Willke and S.~Gossler for advice on laser
noise, N.~Mavalvala and C.~Wipf for comments that helped us to
clarify the message of this paper, as well as Y.K.~Lau, W.~Kells,
H.J.~Kimble and H.~Mabuchi for comments on our manuscript. This
work has been supported by the Alexander~von~Humboldt Foundation's
Sofja~Kovalevskaja programme and by the DFG through the EGC
programme and the SFB No. 407.
\end{acknowledgments}

\end{document}